\documentclass[12pt]{iopart}
\usepackage{iopams}  
\usepackage{graphicx}  
\usepackage{color}  
\usepackage{epsf}
\usepackage{bm}
\usepackage{url}

\newcommand{\KMS}{\rm km\,s^{-1}}
\newcommand{\vz}{v_\|}

\begin{document}

\title[Modeling Gravitational Recoil]
{Modeling Gravitational Recoil Using Numerical Relativity}

\author{Yosef Zlochower, Manuela Campanelli, Carlos O. Lousto}

\address{Center for Computational Relativity and Gravitation, 
School of Mathematical Sciences, 
Rochester Institute of Technology, Rochester, New York 14623, USA}
\ead{}
\begin{abstract}
We review the developments in 
modeling gravitational recoil from merging
black-hole binaries and introduce a new set of 20 simulations
to test our previously proposed
empirical formula for the recoil. The configurations
are chosen to represent generic binaries with unequal masses
and precessing spins. Results of 
these simulations indicate that the recoil formula is
accurate to within a few km/s in the similar mass-ratio
regime for the out-of-plane recoil. 
\end{abstract}

\submitto{\CQG}
\maketitle

\section{Introduction}

The field of Numerical Relativity (NR) has progressed at a remarkable
pace since the breakthroughs of 2005~\cite{Pretorius:2005gq,
Campanelli:2005dd, Baker:2005vv} with the first successful fully
non-linear dynamical numerical simulation of the inspiral, merger, and
ringdown of an orbiting black-hole binary (BHB) system.  In
particular, the `moving-punctures' approach, developed independently
by the NR groups at NASA/GSFC and at RIT, has now become the most
widely used method in the field and was successfully applied 
to evolve generic BHBs.  This approach regularizes a singular
term in space-time metric and allows the black holes (BHs) to 
move across the
computational domain. Previous methods used special coordinate
conditions that kept the BHs fixed in space, which introduced
severe coordinate distortions that caused orbiting-BHB
simulations to crash. Recently, the generalized harmonic approach
method, first developed by Pretorius~\cite{Pretorius:2005gq}, has also
been successfully applied to accurately evolve generic BHBs for tens
of orbits with the use of pseudospectral codes~\cite{Scheel:2008rj,
Szilagyi:2009qz}.

Since then, BHB physics has rapidly matured into a critical tool for
gravitational wave (GW) data analysis and astrophysics.  Recent
developments include: studies of the orbital dynamics of spinning
BHBs~\cite{Campanelli:2006uy, Campanelli:2006fg, Campanelli:2006fy,
Herrmann:2007ex, Marronetti:2007ya, Marronetti:2007wz, Berti:2007fi},
calculations of recoil velocities from the merger of unequal mass
BHBs~\cite{Herrmann:2006ks, Baker:2006vn, Gonzalez:2006md}, and the
surprising discovery that very large recoils can be 
acquired by the remnant of the merger of two spinning BHs
~\cite{Herrmann:2007ac,
Campanelli:2007ew, Campanelli:2007cga, Lousto:2008dn, Pollney:2007ss,
Gonzalez:2007hi, Brugmann:2007zj, Choi:2007eu, Baker:2007gi,
Schnittman:2007ij, Baker:2008md, Healy:2008js, Herrmann:2007zz,
Herrmann:2007ex, Tichy:2007hk, Koppitz:2007ev, Miller:2008en},
empirical models relating the final mass and spin of 
the remnant with the spins of the individual BHs
~\cite{Boyle:2007sz, Boyle:2007ru, Buonanno:2007sv, Tichy:2008du,
Kesden:2008ga, Barausse:2009uz, Rezzolla:2008sd, Lousto:2009mf},
comparisons of waveforms and orbital dynamics of  
BHB inspirals with post-Newtonian (PN) 
predictions~\cite{Buonanno:2006ui, Baker:2006ha, Pan:2007nw,
Buonanno:2007pf, Hannam:2007ik, Hannam:2007wf, Gopakumar:2007vh,
Hinder:2008kv}, and simulations reaching mass ratios \cite{Lousto:2010ut}
$q=1/100$.

\subsection{Kicks and Super-kicks: A brief history}

The first in-depth modeling of the recoil from the merger of
non-spinning asymmetric BHBs was done in Ref.~\cite{Gonzalez:2006md},
where it was shown that the maximum recoil is limited to $\approx 175\
\KMS$. Soon after, other groups showed that the maximum recoil for
spinning binaries is much larger. In
Ref.~\cite{Herrmann:2007ac}~and~\cite{Koppitz:2007ev} (which were
released within a few days of each other), it was shown that the
maximum recoil for a spinning binary with one BH spin aligned with the
orbital angular momentum and other anti-aligned id $475\ \KMS$. Within
a day of the initial release of the preprint for
Ref.~\cite{Herrmann:2007ac}, our group released a
preprint~\cite{Campanelli:2007ew} for more generic spin and mass
configurations, showing that the recoil is considerably larger if the
spins are anti-aligned and pointing in the orbital plane.
In~\cite{Campanelli:2007ew} we measured recoil velocities beyond
the maximum predicted for the configurations
in~\cite{Herrmann:2007ac, Koppitz:2007ev}.

 An initial analysis of the
results in~\cite{Campanelli:2007ew} indicated that the maximum
velocity exceeded $1300\ \KMS$ for spins lying in the orbital plane.
 This initial estimate was based on
two `generic' configurations. A subsequent analysis that incorporated
the angular dependence of the projection of the spin on the orbital 
plane, showed that the
maximum recoils is $\sim4000\ \KMS$~\cite{Campanelli:2007cg}.
Based on our conclusions in the preprint of~\cite{Campanelli:2007ew},
the group at Jena performed a set of two simulations in this
`maximum-kick' configuration and measured recoils of around $2600\ \KMS$. Our
modeling showed that this recoil velocity actually depends
sinusoidally on the angle that the spins make with the infall
direction. By evolving a set of configurations with spins at different
initial angles in the orbital plane, we found that the recoil
reaches a maximum of $\sim 4000\ \KMS$ 
\cite{Campanelli:2007cga,Lousto:2010xk}. In
Ref.~\cite{Campanelli:2007cg}, as part of our analysis, we
proposed an empirical formula for the recoil based on post-Newtonian
expressions for the radiated linear momentum. The formula correctly
predicts the sinusoidal dependence of the `maximum-kick' recoil.

In Ref.~\cite{Baker:2008md} the recoil for unequal-mass, spinning
binaries, with total spin equal to zero and individual spins lying in
the initial orbital plane, was measured. These  $S=0$ configurations
are preserved under numerical evolution and lead to minimal precession
of the orbital pane. Interestingly, they found that the recoil scales
with the cube of the mass ratio $q$, rather than the expected 
$q^2$ seen in post-Newtonian expressions for the recoil. In a subsequent
study~\cite{Lousto:2008dn}, our group found that the recoil scales as
$q^2$ for the more astrophysically important precessing case.

\subsection{Phenomenological modeling of the recoil}

In~\cite{Campanelli:2007ew} we introduced an empirical
formula for the recoil, augmented in~\cite{Lousto:2009mf}
which has the form.
\begin{eqnarray}\label{eq:Pempirical}
\vec{V}_{\rm recoil}&&(q,\vec\alpha)=v_m\,\hat{e}_1+
v_\perp(\cos\xi\,\hat{e}_1+\sin\xi\,\hat{e}_2)+v_\|\,\hat{n}_\|,\nonumber\\
&&v_m=A\frac{\eta^2(1-q)}{(1+q)}\left[1+B\,\eta\right],\nonumber\\
&&v_\perp=H\frac{\eta^2}{(1+q)}\left[
(1+B_H\,\eta)\,(\alpha_2^\|-q\alpha_1^\|)
+\,H_S\,\frac{(1-q)}{(1+q)^2}\,(\alpha_2^\|+q^2\alpha_1^\|)\right],\nonumber\\
&&v_\|=K\frac{\eta^2}{(1+q)}\Bigg[
(1+B_K\,\eta)
\left|\alpha_2^\perp-q\alpha_1^\perp\right|
\cos(\Theta_\Delta-\Theta_0)\nonumber\\
&&\quad+\,K_S\,\frac{(1-q)}{(1+q)^2}\,\left|\alpha_2^\perp+q^2\alpha_1^\perp\right|
\cos(\Theta_S-\Theta_1)\Bigg],
\end{eqnarray}
where $\eta=q/(1+q)^2$, 
 the index $\perp$ and $\|$ refer to
perpendicular and parallel to the orbital angular momentum
respectively,
$\hat{e}_1,\hat{e}_2$ are
orthogonal unit vectors in the orbital plane, and $\xi$ measures the
angle between the unequal mass and spin contribution to the recoil
velocity in the orbital plane. The constants $H_S$ and $K_S$ can
be
determined from new generic BHB simulations as the data become
 available. The angles, $\Theta_\Delta$ and $\Theta_S$, are
the angles
between the in-plane component of $\vec \Delta = M (\vec S_2/m_2 -
\vec
S_1/m_1)$ or $\vec S=\vec S_1+\vec S_2$
and the infall direction at merger. Phases $\Theta_0$ and $\Theta_1$
depend
on the initial separation of the holes for quasicircular orbits.
A crucial observation is that the dominant contribution to the recoil
is generated near the time of formation of the common horizon of the
merging BHs (See, for instance Fig. 6 in
~\cite{Lousto:2007db}).  The formula (\ref{eq:Pempirical}) above
describing the recoil applies at this moment (or averaged coefficients
around this maximum generation of recoil), and has proven to represent
 the distribution of velocities with sufficient accuracy for
astrophysical
applications. 
Recently, in Ref.~\cite{vanMeter:2010md} a very similar formula
was proposed.

The most recent published estimates for the above parameters can be found in
\cite{Lousto:2008dn} and references therein. The current best
estimates are: $A = 1.2\times 10^{4}\
\KMS$, $B = -0.93$, $H = (6.9\pm0.5)\times 10^{3}\ \KMS$,
$K=(6.0\pm0.1)\times 10^4\ \KMS$, and $\xi \sim 145^\circ$.

Note that $\vz$ is maximized when $q=1$ (i.e.\ equal masses),
the in-plane spins are maximal ($\alpha_1 = \alpha_2=1$),
and the two spins are anti-aligned.

In this paper, we use a new set of simulations of precessing binaries
to test the
recoil formula and obtain a new estimate for $K_S$.

\section{Numerical Techniques}

To compute the numerical initial data, we use the puncture
approach~\cite{Brandt97b} along with the {\sc
TwoPunctures}~\cite{Ansorg:2004ds} thorn.  In this approach the
3-metric on the initial slice has the form $\gamma_{a b} = (\psi_{BL}
+ u)^4 \delta_{a b}$, where $\psi_{BL}$ is the Brill-Lindquist
conformal factor, $\delta_{ab}$ is the Euclidean metric, and $u$ is
(at least) $C^2$ on the punctures.  The Brill-Lindquist conformal
factor is given by $ \psi_{BL} = 1 + \sum_{i=1}^n m_{i}^p / (2 |\vec r
- \vec r_i|), $ where $n$ is the total number of `punctures',
$m_{i}^p$ is the mass parameter of puncture $i$ ($m_{i}^p$ is {\em
not} the horizon mass associated with puncture $i$), and $\vec r_i$ is
the coordinate location of puncture $i$.  We evolve these
BHB data-sets using the {\sc
LazEv}~\cite{Zlochower:2005bj} implementation of the moving puncture
approach~\cite{Campanelli:2005dd,Baker:2005vv} with the conformal
factor $W=\sqrt{\chi}=\exp(-2\phi)$ suggested by~\cite{Marronetti:2007wz}
For the runs presented here
we use centered, eighth-order finite differencing in
space~\cite{Lousto:2007rj} and an RK4 time integrator (note that we do
not upwind the advection terms).

We use the Carpet~\cite{Schnetter-etal-03b} mesh refinement driver to
provide a `moving boxes' style mesh refinement. In this approach
refined grids of fixed size are arranged about the coordinate centers
of both holes.  The Carpet code then moves these fine grids about the
computational domain by following the trajectories of the two BHs.

We use {\sc AHFinderDirect}~\cite{Thornburg2003:AH-finding} to locate
apparent horizons.  We measure the magnitude of the horizon spin using
the Isolated Horizon algorithm detailed in~\cite{Dreyer02a}. This
algorithm is based on finding an approximate rotational Killing vector
(i.e.\ an approximate rotational symmetry) on the horizon $\varphi^a$. Given
this approximate Killing vector $\varphi^a$, the spin magnitude is
\begin{equation}
 \label{isolatedspin} S_{[\varphi]} =
 \frac{1}{8\pi}\int_{AH}(\varphi^aR^bK_{ab})d^2V,
\end{equation}
where $K_{ab}$ is the extrinsic curvature of the 3D-slice, $d^2V$ is
the natural volume element intrinsic to the horizon, and $R^a$ is the
outward pointing unit vector normal to the horizon on the 3D-slice.
We measure the direction of the spin by finding the coordinate line
joining the poles of this Killing vector field using the technique
introduced in~\cite{Campanelli:2006fy}.  Our algorithm for finding the
poles of the Killing vector field has an accuracy of $\sim 2^\circ$
(see~\cite{Campanelli:2006fy} for details). Note that once we have the
horizon spin, we can calculate the horizon mass via the Christodoulou
formula
${m^H} = \sqrt{m_{\rm irr}^2 +
 S^2/(4 m_{\rm irr}^2)},
$
where $m_{\rm irr} = \sqrt{A/(16 \pi)}$ and $A$ is the surface area of
the horizon.
We measure radiated energy, linear momentum, and angular momentum, in
terms of $\psi_4$, using the formulae provided in
Refs.~\cite{Campanelli99,Lousto:2007mh}. However, rather than using
the full $\psi_4$, we decompose it into $\ell$ and $m$ modes and solve
for the radiated linear momentum, dropping terms with $\ell \geq 5$.
The formulae in Refs.~\cite{Campanelli99,Lousto:2007mh} are valid at
$r=\infty$. 
We obtain accurate values for these quantities by
solving for them on spheres of finite radius (typically $r/M=50, 60,
\cdots, 100$), fitting the results to a polynomial dependence in
$l=1/r$, and extrapolating to
$l=0$~\cite{Baker:2005vv,Campanelli:2006gf,Hannam:2007ik,Boyle:2007ft}. Each quantity $Q$ has the radial
dependence $Q=Q_0 + l Q_1 + {\cal O}(l^2)$, where $Q_0$ is the
asymptotic value (the ${\cal O}(l)$ error arises from the ${\cal
O}(l)$ error in $r\, \psi_4$). We perform both linear and quadratic
fits of $Q$ versus $l$, and take $Q_0$ from the quadratic fit as the
final value with the differences between the linear and extrapolated
$Q_0$ as a measure of the error in the extrapolations.
 We found that extrapolating
the waveform itself to $r=\infty$ introduced  phase errors due to
uncertainties in the areal radius of the observers, as well as
numerical noise. Thus when comparing Perturbative to numerical waveforms, we use
the waveform extracted at $r=100M$. 

In order to model the recoil as a function of the orientation and
magnitudes of the spins, we use the techniques introduced
in~\cite{Lousto:2008dn}
to locate the approximate orbital plane at merger and 3D rotation such
that  infall directions are the same for each
simulation. Briefly, this technique uses three points on the
trajectories, given by fiducial choices of the BH separations, to define
the orbital plane and preferred orientation.

\section{Simulations}

\begin{table}[ht]\label{table:ID}
\begin{center}
\begin{tabular}{lccccccccccc}
\hline
Config   & $x_1/M$ & $x_2/M$  & $P/M$    & $m^p_1$ & $m^p_2$ &
$S_x/M^2$ & $S_y/
M^2$ \\ 
\hline
Q33TH000 & 4.882446 & -1.607923 & 0.101163 & 0.171173 &
   0.723529 & 0 & 0.045983 \\ 
RQ33TH000 & 4.885558 & -1.609045 & 0.101153 & 0.171170 &
   0.723523 & 0 & 0.045982 \\ 
Q50TH000 & 4.360493 & -2.163334 & 0.119252 & 0.230648 &
  0.618813 & 0 &  0.082098\\
\hline
\end{tabular}
\caption {Initial data parameters for the quasi-circular
configurations. The punctures are
located
at $\vec r_1 = (x_1,0,0)$ and $\vec r_2 = (x_2,0,0)$, with momenta
$P=\pm (0, P,0)$, spins $\vec S_1 = (S_x, S_y, 0)$,
 $\vec S_2 = - q\,\vec S_1$, mass parameters
$m^p$. The configuration are denoted by QXXXTHYYY where XXX
gives the mass ratio (0.33, 0.50) and
YYY
gives the angle in degrees
between the initial spin direction and the $y$-axis. In all cases the
initial orbital period is $M \omega = 0.05$ and the spin of the
smaller
BH is $\alpha=0.72$. Initial data parameters
for the Q50TH000 and Q33TH000 configurations are given. The
remaining configurations are obtained by rotating the spin
directions, keeping all other parameters the same. For the
RQ33THxxx configurations, $\vec S_2$ is rotated by $90^\circ$ with
respect to the corresponding Q33THxxx configuration.}
\end{center}
\end{table}

We evolve a set of configuration that initially have $\Delta=0$, as
well as a set of configuration with one spin rotated by $90^\circ$ for
mass ratios $q=1/3$ and $q=1/2$. The initial data parameters
are summarized in Table~\ref{table:ID}. Note that the $\Delta=0$
configurations are unstable in the sense that the system
quickly evolved towards a nontrivial $\Delta$. In particular,
at merger, where most of the recoil asymmetry is generated, the 
binary reaches a generic configuration regarding the spin orientations
given the strong differential precession of each BH spins.

\section{Results}

In Table~\ref{table:raw_kick} we summarize the results of the
simulations. The table shows the radiated energy and recoil (prior to
any rotation). Note that these results can be used in additional fits
of the final remnant BH formulae for the mass, spin, and
recoil velocity of the remnant, as was done in
Ref.~\cite{Lousto:2009mf} using the then currently available results
in the literature.

\begin{table}[ht]
\begin{center}
\begin{tabular}{lcccc}
\hline
Config   & $100(\delta E/M)$ & $V_x$ & $ V_y$ & $V_z$\\
\hline
Q50TH000 & $2.858\pm0.018$ & $76.89\pm2.47$ & $234.40\pm3.12$ &
  $167.95\pm3.29$\\
Q50TH045 & $2.817\pm0.017$ & $-40.47\pm131.03$ & $10.46\pm99.56$
& $-228.49\pm5.76$ \\
Q50TH090 & $2.778\pm0.016$ & $189.23\pm 0.77$ & $97.60\pm
0.34$ & $-543.78\pm 2.29$ \\
Q50TH130 & $2.795\pm0.017$ & $306.54\pm1.23$ & $319.10\pm0.02$ &
$-762.32\pm0.29$ \\
Q50TH180 & $2.858\pm0.018$ & $78.03\pm2.66$ & $235.61\pm3.24$ &
$-167.47\pm3.37$ \\
Q50TH210 & $2.831\pm0.017$ & $43.42\pm1.32$ & $108.60\pm2.94$ &
$131.29\pm3.42$ \\
Q50TH315 & $2.831\pm0.017$ & $295.55\pm1.26$ & $341.01\pm0.78$ &
$741.05\pm0.66$\\
\hline
Q33TH000 & $1.90243\pm0.012$ & $119.18\pm0.36$ & $212.92\pm2.78$ & 
 $144.19 1.08 $\\
Q33TH045 & $1.884\pm0.012$ & $67.29\pm0.59$ & $132.08\pm2.51$ &
$-134.74\pm0.94$\\
Q33TH090 & $1.878\pm0.012$ & $02.92\pm1.87$ & $132.08\pm1.51$ &
$-337.04\pm0.88$\\
Q33TH130 & $1.897\pm0.012$ & $210.61\pm1.34$ & $235.72\pm0.71$ &
$-497.12\pm 1.34$\\
Q33TH180 & $1.902\pm0.012$ & $119.21\pm0.36$ & $212.83\pm2.78$ &
$-144.04\pm1.08$\\
Q33TH210 & $1.889\pm0.012$ & $73.21\pm0.11$ & $150.46\pm2.73$ &
$59.42\pm1.02$\\
Q33TH260 & $1.878\pm0.012$ & $73.45\pm1.79$ & $127.45\pm1.58$ &
$272.87\pm1.14$\\
Q33TH315 & $1.901\pm0.013$ & $210.14\pm1.45$ & $249.05\pm0.87$ &
$489.34\pm0.76$\\
\hline
RQ33TH000 & $1.885\pm0.013$ & $229.17\pm1.07$ & $132.08\pm1.81$ &
$539.21\pm1.66$\\
RQ33TH090 & $1.865\pm0.012$ & $45.62\pm0.11$ & $162.27\pm1.00$ &
$-74.423\pm0.08$\\
RQ33TH130 & $1.860\pm0.012$ & $106.45\pm1.16$ & $82.27\pm0.45$ &
$-427.92\pm1.43$\\
RQ33TH210 & $1.886\pm0.012$ & $182.61\pm0.67$ & $188.75\pm2.00$ &
$-361.19\pm$1.04\\
RQ33TH315 & $1.861\pm0.012$ & $121.99\pm1.42$ & $79.03\pm0.45$ &
$462.96\pm1.68$\\
\hline
\end{tabular}
\caption{The radiated energy and 
recoil velocities for each configuration.
Note that some of the error estimates, which are based on the
differences between a linear and quadratic
extrapolation in $l=1/r$ of the observer location, are very small.
This indicates that the
differences between the extrapolation can underestimate
the true error. All quantities are given in the coordinate system used
by the code (i.e.\ the untransformed system). }
\label{table:raw_kick}
\end{center}
\end{table}

Table \ref{table:raw_ang} gives the components of the radiated angular
momentum in the original $x,y,z$ frame (that of the initial data) using
the Cartesian decomposition as in Ref.\ \cite{Lousto:2007mh}.

\begin{table}[ht]
\begin{center}
\begin{tabular}{lcccc}
\hline
Config   & $\delta J_{x}$ & $\delta J_{y}$ & $\delta J_{z}$ \\
\hline
Q50TH000 & $0.02454 \pm 0.00056$ & $0.05954 \pm 0.00012$ & $0.18040 \pm 0.00200$ \\
Q50TH045 & $-0.0277 \pm 0.0041$ & $0.0572 \pm 0.0029$ & $0.1808 \pm 0.0017$ \\
Q50TH090 & $-0.05995 \pm 0.00019$ & $ 0.02354 \pm 0.00079$ & $0.1808 \pm 0.0015$ \\
Q50TH130 & $-0.06155 \pm 0.00029$ & $-0.02016 \pm 0.00047$ & $0.1801 \pm 0.00164$ \\
Q50TH180 & $-0.02454 \pm 0.00056$ & $-0.05954 \pm 0.00012$ & $0.1804 \pm 0.0020$ \\
Q50TH210 & $0.00857 \pm 0.00035$ & $-0.06326 \pm 0.00059$ & $0.1805 \pm 0.0018$ \\
Q50TH315 & $0.05981 \pm 0.00026$ & $0.02545 \pm 0.00046$ & $0.1802 \pm 0.0017$ \\
\hline
Q33TH000 & $0.0180 \pm 0.00011$ & $0.03134 \pm 0.00022$ & $0.1309 \pm 0.0012$ \\
Q33TH045 & $-0.00930 \pm 0.00008$ & $0.03504 \pm 0.00010$ & $0.1312 \pm 0.0012$ \\
Q33TH090 & $-0.03151 \pm 0.00002$ & $0.01835 \pm 0.00013$ & $0.13262 \pm 0.00081$ \\
Q33TH130 & $-0.03524 \pm 0.00047$ & $-0.00678 \pm 0.00002$ & $0.1325 \pm 0.0010$ \\
Q33TH180 & $-0.01801 \pm 0.00011$ & $-0.03134 \pm 0.00022$ & $0.1309 \pm 0.0012$ \\
Q33TH210 & $-0.00013 \pm 0.00002$ & $-0.03608 \pm 0.00022$ & $0.1309 \pm 0.0012$ \\
Q33TH260 & $0.02797 \pm 0.00003$ & $-0.02366 \pm 0.00014$ & $0.13254 \pm 0.00084$ \\
Q33TH315 & $0.03462 \pm 0.00041$ & $0.00994 \pm 0.00002$ & $0.1325 \pm 0.0011$ \\
\hline
RQ33TH000 & $0.021251 \pm 0.000001$ & $-0.00413 \pm 0.00003$ & $0.1348 \pm 0.0012$ \\
RQ33TH090 & $0.00462 \pm 0.00017$ & $0.02225 \pm 0.00001$ & $0.13479 \pm 0.00091$ \\
RQ33TH130 & $-0.00981 \pm 0.00001$ & $0.01968 \pm 0.00017$ & $0.1346 \pm 0.0010$ \\
RQ33TH210 & $-0.02076 \pm 0.00014$ & $-0.00734 \pm 0.00008$ & $0.1345 \pm 0.0012$ \\
RQ33TH315 & $0.01143 \pm 0.00001$ & $-0.01857 \pm 0.00022$ & $0.1346 \pm 0.0011$ \\
\hline
\end{tabular}
\caption{The radiated angular momentum.
Note that some of the error estimates, which are based on the
differences between a linear and quadratic
extrapolation in $l=1/r$ of the observer location, are very small.
This indicates that the
differences between the extrapolation can underestimate
the true error. All quantities are given in the coordinate system used
by the code (i.e.\ the untransformed system). }
\label{table:raw_ang}
\end{center}
\end{table}

\begin{figure}
\begin{center}
 \includegraphics[width=4in]{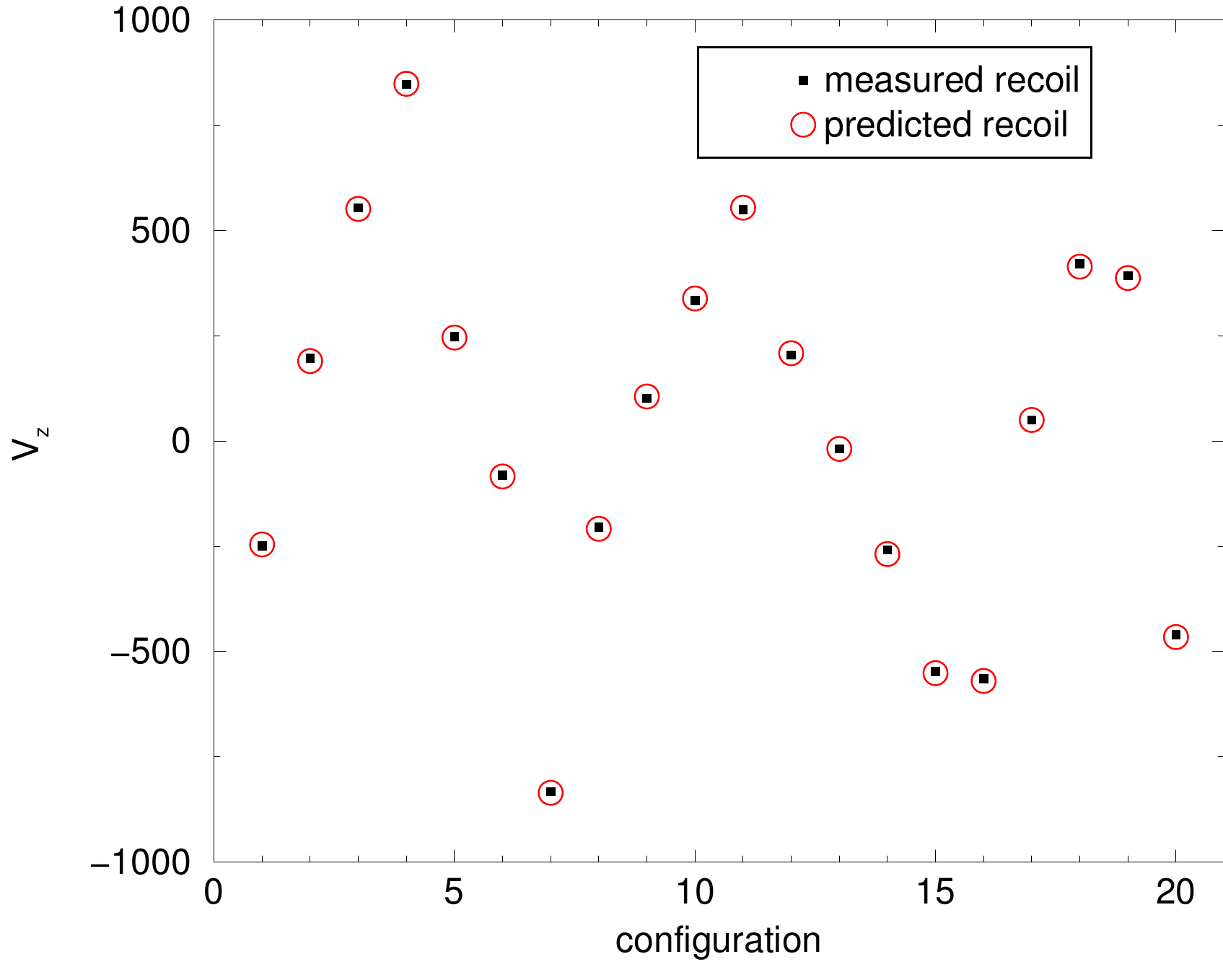}
  \caption{The predicted and measured out-of-plane recoil 
for each configuration in Table~\ref{table:rotated_kick_and_fit}.
The configurations are labeled 1-20 starting from Q50TH000
as ordered in Table \ref{table:rotated_kick_and_fit}.}
\label{fig:kick_fit}
\end{center}
\end{figure}

In Table~\ref{table:rotated_kick_and_fit} we
give the recoil velocities in a frame rotated such that the
orbital plane coincides with the $xy$ plane and the
infall direction in this plane is fixed. In order to fit
the data, we chose $K=5.9\times10^4$ based on previous work 
\cite{Campanelli:2007ew,Lousto:2008dn,Lousto:2010xk}
and then fit the $Q50$ simulations
for $K_{S}$, $\Theta_0$, and $\Theta_1$. We found
$K_s=-4.2\pm1.8$. While, $K$ and $K_S$ are fixed,
$\Theta_0$ and $\Theta_1$ depend on the
configuration. To obtain these angles we fit the
the recoil formula for the Q33Txxx and RQ33Txxx configurations
separately. We then compared the predicted recoil for each
configuration with the measured recoil.
The results
are summarized in Table~\ref{table:rotated_kick_and_fit}. Note that the
errors are typically less than $6\ \KMS$. The largest relative errors
are $\sim 3.3\%$, with most errors lying between $\sim 1\%$ and
$\sim2\%$, rendering the empirical formula accurate
for most astrophysical applications. We plot the measured out-of-plane recoil
and prediction in Fig.~\ref{fig:kick_fit}. An attempt to fit the
in-plane recoil produced errors of the order of $50\KMS$. This can be
traced to three main sources the errors. The  uncertainty in how the the 
recoils produced by unequal masses and
out-of plane spins contribute to the total in-plane recoil, the
error in the measurement in the out-of-plane spine due to errors in
measuring the orientation of the orbital plane at merger,
and, as pointed out in~\cite{Lousto:2008dn},
 effects of precession on the in-plane
recoil. We note that the dominant out-of-plane recoil is well modeled.

\begin{table}[ht]
\begin{center}
\begin{tabular}{lccccc}
\hline
Config   & $V_x$ & $ V_y$ & $V_z$ & $V_z$(predict) & error \\
\hline
Q50TH000 & 17.47 & -164.42 & -248.44 & -245.851 & 2.591 \\
Q50TH045 & -71.85 & -100.95 & 196.47 & 190.25 & -6.222 \\
Q50TH090 & 44.52 & -180.82 & 553.49 & 551.759 & -1.73 \\
Q50TH130 & 31.47 & -242.82 & 846.74 & 848.283 & 1.540 \\
Q50TH180 & 18.34 & -165.91 & 248.56 & 245.987 & -2.575 \\
Q50TH210 & 26.93 & -153.45 & -81.52 & -84.3437 & -2.823 \\
Q50TH315 & 28.58 & -242.02 & -832.71 & -834.713 & -1.999 \\
\hline
Q33TH000 & 117.61 & -157.99 & -203.81 & -208.612 & -3.311 \\
Q33TH045 & 110.76 & -132.11 & 102.02 & 105.811 & 1.937 \\
Q33TH090 & 122.60 & -120.81 & 334.67 & 337.689 & -0.893 \\
Q33TH130 & 144.38 & -155.39 & 549.60 & 553.834 & -2.540 \\
Q33TH180 & 117.60 & -157.99 & 203.63 & 208.286 & 3.140 \\
Q33TH210 & 109.72 & -138.43 & -18.17 & -18.7138 & 0.501 \\
Q33TH260 & 118.62 & -122.29 & -258.97 & -268.242 & -6.040 \\
Q33TH315 & 144.43 & -160.93 & -546.70 & -551.418 & 2.112 \\
\hline
RQ33TH000 & 168.76 & -118.44 & -564.09 & -569.559 & -5.469 \\
RQ33TH090 & 116.46 & -133.88 & 49.68 & 49.8794 & 0.203 \\
RQ33TH130 & 121.09 & -92.33 & 421.93 & 413.897 & -8.033 \\
RQ33TH210 & 155.20 & -147.31 & 391.98 & 386.733 & -5.246 \\
RQ33TH315 & 125.67 & -89.20 & -460.12 & -464.877 & -4.757 \\
\hline
\end{tabular}
\caption{The recoil velocities after rotation and a comparison of the
out-of-plane component of the recoil (in
coordinates aligned with the orbital plane at merger) with the
predictions of the empirical formula with coefficients $K=5.9\times
10^4$ and $K_S = -4.25401$.}
\label{table:rotated_kick_and_fit}
\end{center}
\end{table}

\section{Conclusion}

We report here on a new set of generic simulations, with no symmetries
and different choices of the binary's mass ratio and spin direction
and magnitudes. We compute the radiation emitted by this binaries, and
in particular focus on the magnitude and direction of the final recoil
of the merged BH. While this set of runs can be used for
fitting to additional subleading terms in the empirical formulae 
for the remnant BH recoil, we choose to
use them to test the previously fitted values, extended to include
effects linear in the spin, as accurate determinations of these
parameters will require many more simulations.
In this paper we showed that the empirical recoil formula provides 
accurate predictions for the recoil velocity from BHB mergers
for the new sets of configurations. These configurations
have fewer symmetries than previous comparisons and can be considered
of generic nature regarding spin orientations and intermediate mass
ratios and spin magnitudes. This shows that our empirical formula
(\ref{eq:Pempirical}) can be used as a first approximation for astrophysical
studies of statistical nature, as we did in Ref.\ \cite{Lousto:2009ka} and also
used for realistic recoil magnitudes and direction when modeling
the observational effects of recoiling BHs in a gaseous
environment such as accretion disks \cite{Anderson:2009fa}.

\medskip
\noindent
{\it Acknowledgments:}
We gratefully acknowledge NSF for financial support from grants
PHY-0722315, PHY-0653303, PHY-0714388, PHY-0722703, DMS-0820923,
PHY-0969855, PHY-0903782, PHY-0929114 and CDI-1028087;
and NASA for financial support from grants NASA
07-ATFP07-0158 and HST-AR-11763.  Computational resources were
provided by Ranger cluster at TACC (Teragrid allocation TG-PHY060027N)
and by NewHorizons at RIT.

\section*{References}

\bibliographystyle{unsrt}
\bibliography{../../../Bibtex/references}

\begin{thebibliography}{10}

\bibitem{Pretorius:2005gq}
Frans Pretorius.
\newblock Evolution of binary black hole spacetimes.
\newblock {\em Phys. Rev. Lett.}, 95:121101, 2005.

\bibitem{Campanelli:2005dd}
Manuela Campanelli, C.~O. Lousto, P.~Marronetti, and Y.~Zlochower.
\newblock Accurate evolutions of orbiting black-hole binaries without excision.
\newblock {\em Phys. Rev. Lett.}, 96:111101, 2006.

\bibitem{Baker:2005vv}
John~G. Baker, Joan Centrella, Dae-Il Choi, Michael Koppitz, and James van
  Meter.
\newblock Gravitational wave extraction from an inspiraling configuration of
  merging black holes.
\newblock {\em Phys. Rev. Lett.}, 96:111102, 2006.

\bibitem{Scheel:2008rj}
Mark~A. Scheel et~al.
\newblock {High-accuracy waveforms for binary black hole inspiral, merger, and
  ringdown}.
\newblock {\em Phys. Rev.}, D79:024003, 2009.

\bibitem{Szilagyi:2009qz}
Bela Szilagyi, Lee Lindblom, and Mark~A. Scheel.
\newblock {Simulations of Binary Black Hole Mergers Using Spectral Methods}.
\newblock {\em Phys. Rev.}, D80:124010, 2009.

\bibitem{Campanelli:2006uy}
Manuela Campanelli, C.~O. Lousto, and Y.~Zlochower.
\newblock Spinning-black-hole binaries: The orbital hang up.
\newblock {\em Phys. Rev. D}, 74:041501(R), 2006.

\bibitem{Campanelli:2006fg}
Manuela Campanelli, C.~O. Lousto, and Yosef Zlochower.
\newblock Spin-orbit interactions in black-hole binaries.
\newblock {\em Phys. Rev. D}, 74:084023, 2006.

\bibitem{Campanelli:2006fy}
Manuela Campanelli, Carlos~O. Lousto, Yosef Zlochower, Badri Krishnan, and
  David Merritt.
\newblock Spin flips and precession in black-hole-binary mergers.
\newblock {\em Phys. Rev.}, D75:064030, 2007.

\bibitem{Herrmann:2007ex}
Frank Herrmann, Ian Hinder, Deirdre~M. Shoemaker, Pablo Laguna, and Richard~A.
  Matzner.
\newblock {Binary Black Holes: Spin Dynamics and Gravitational Recoil}.
\newblock {\em Phys. Rev.}, D76:084032, 2007.

\bibitem{Marronetti:2007ya}
Pedro Marronetti et~al.
\newblock Binary black holes on a budget: Simulations using workstations.
\newblock {\em Class. Quant. Grav.}, 24:S43--S58, 2007.

\bibitem{Marronetti:2007wz}
Pedro Marronetti, Wolfgang Tichy, Bernd Brugmann, Jose Gonzalez, and Ulrich
  Sperhake.
\newblock {High-spin binary black hole mergers}.
\newblock {\em Phys. Rev.}, D77:064010, 2008.

\bibitem{Berti:2007fi}
Emanuele Berti et~al.
\newblock Inspiral, merger and ringdown of unequal mass black hole binaries: A
  multipolar analysis.
\newblock {\em Phys. Rev.}, D76:064034, 2007.

\bibitem{Herrmann:2006ks}
F.~Herrmann, D.~Shoemaker, and P.~Laguna.
\newblock Unequal-mass binary black hole inspirals.
\newblock {\em AIP Conf.}, 873:89--93, 2006.

\bibitem{Baker:2006vn}
John~G. Baker et~al.
\newblock Getting a kick out of numerical relativity.
\newblock {\em Astrophys. J.}, 653:L93--L96, 2006.

\bibitem{Gonzalez:2006md}
Jose~A. Gonz\'alez, Ulrich Sperhake, Bernd Brugmann, Mark Hannam, and Sascha
  Husa.
\newblock Total recoil: the maximum kick from nonspinning black-hole binary
  inspiral.
\newblock {\em Phys. Rev. Lett.}, 98:091101, 2007.

\bibitem{Herrmann:2007ac}
Frank Herrmann, Ian Hinder, Deirdre Shoemaker, Pablo Laguna, and Richard~A.
  Matzner.
\newblock Gravitational recoil from spinning binary black hole mergers.
\newblock {\em Astrophys. J.}, 661:430--436, 2007.

\bibitem{Campanelli:2007ew}
Manuela Campanelli, Carlos~O. Lousto, Yosef Zlochower, and David Merritt.
\newblock Large merger recoils and spin flips from generic black-hole binaries.
\newblock {\em Astrophys. J.}, 659:L5--L8, 2007.

\bibitem{Campanelli:2007cga}
Manuela Campanelli, Carlos~O. Lousto, Yosef Zlochower, and David Merritt.
\newblock Maximum gravitational recoil.
\newblock {\em Phys. Rev. Lett.}, 98:231102, 2007.

\bibitem{Lousto:2008dn}
Carlos~O. Lousto and Yosef Zlochower.
\newblock {Modeling gravitational recoil from precessing highly- spinning
  unequal-mass black-hole binaries}.
\newblock {\em Phys. Rev. D}, 79:064018, 2009.

\bibitem{Pollney:2007ss}
Denis Pollney et~al.
\newblock {Recoil velocities from equal-mass binary black-hole mergers: a
  systematic investigation of spin-orbit aligned configurations}.
\newblock {\em Phys. Rev.}, D76:124002, 2007.

\bibitem{Gonzalez:2007hi}
J.~A. Gonz\'alez, M.~D. Hannam, U.~Sperhake, B.~Brugmann, and S.~Husa.
\newblock Supermassive kicks for spinning black holes.
\newblock {\em Phys. Rev. Lett.}, 98:231101, 2007.

\bibitem{Brugmann:2007zj}
Bernd Brugmann, Jose~A. Gonzalez, Mark Hannam, Sascha Husa, and Ulrich
  Sperhake.
\newblock {Exploring black hole superkicks}.
\newblock {\em Phys. Rev.}, D77:124047, 2008.

\bibitem{Choi:2007eu}
Dae-Il Choi et~al.
\newblock Recoiling from a kick in the head-on collision of spinning black
  holes.
\newblock {\em Phys. Rev.}, D76:104026, 2007.

\bibitem{Baker:2007gi}
John~G. Baker et~al.
\newblock Modeling kicks from the merger of non-precessing black-hole binaries.
\newblock {\em Astrophys. J.}, 668:1140--1144, 2007.

\bibitem{Schnittman:2007ij}
Jeremy~D. Schnittman et~al.
\newblock {Anatomy of the binary black hole recoil: A multipolar analysis}.
\newblock {\em Phys. Rev.}, D77:044031, 2008.

\bibitem{Baker:2008md}
John~G. Baker et~al.
\newblock {Modeling kicks from the merger of generic black-hole binaries}.
\newblock {\em Astrophys. J.}, 682:L29, 2008.

\bibitem{Healy:2008js}
James Healy et~al.
\newblock {Superkicks in Hyperbolic Encounters of Binary Black Holes}.
\newblock {\em Phys. Rev. Lett.}, 102:041101, 2009.

\bibitem{Herrmann:2007zz}
Frank Herrmann, Ian Hinder, Deirdre Shoemaker, and Pablo Laguna.
\newblock Unequal mass binary black hole plunges and gravitational recoil.
\newblock {\em Class. Quant. Grav.}, 24:S33--S42, 2007.

\bibitem{Tichy:2007hk}
Wolfgang Tichy and Pedro Marronetti.
\newblock Binary black hole mergers: Large kicks for generic spin orientations.
\newblock {\em Phys. Rev.}, D76:061502, 2007.

\bibitem{Koppitz:2007ev}
Michael Koppitz et~al.
\newblock Getting a kick from equal-mass binary black hole mergers.
\newblock {\em Phys. Rev. Lett.}, 99:041102, 2007.

\bibitem{Miller:2008en}
Sarah~H. Miller and R.~A. Matzner.
\newblock {Multipole Analysis of Kicks in Collision of Binary Black Holes}.
\newblock {\em Gen. Rel. Grav.}, 41:525--539, 2009.

\bibitem{Boyle:2007sz}
Latham Boyle, Michael Kesden, and Samaya Nissanke.
\newblock {Binary black hole merger: symmetry and the spin expansion}.
\newblock {\em Phys. Rev. Lett.}, 100:151101, 2008.

\bibitem{Boyle:2007ru}
Latham Boyle and Michael Kesden.
\newblock {The spin expansion for binary black hole merger: new predictions and
  future directions}.
\newblock {\em Phys. Rev.}, D78:024017, 2008.

\bibitem{Buonanno:2007sv}
Alessandra Buonanno, Lawrence~E. Kidder, and Luis Lehner.
\newblock {Estimating the final spin of a binary black hole coalescence}.
\newblock {\em Phys. Rev.}, D77:026004, 2008.

\bibitem{Tichy:2008du}
Wolfgang Tichy and Pedro Marronetti.
\newblock {The final mass and spin of black hole mergers}.
\newblock {\em Phys. Rev.}, D78:081501, 2008.

\bibitem{Kesden:2008ga}
Michael Kesden.
\newblock {Can binary mergers produce maximally spinning black holes?}
\newblock {\em Phys. Rev.}, D78:084030, 2008.

\bibitem{Barausse:2009uz}
Enrico Barausse and Luciano Rezzolla.
\newblock {Predicting the direction of the final spin from the coalescence of
  two black holes}.
\newblock {\em Astrophys. J. Lett.}, 704:L40--L44, 2009.

\bibitem{Rezzolla:2008sd}
Luciano Rezzolla.
\newblock {Modelling the final state from binary black-hole coalescences}.
\newblock {\em Class. Quant. Grav.}, 26:094023, 2009.

\bibitem{Lousto:2009mf}
Carlos~O. Lousto, Manuela Campanelli, Yosef Zlochower, and Hiroyuki Nakano.
\newblock {Remnant Masses, Spins and Recoils from the Merger of Generic
  Black-Hole Binaries}.
\newblock {\em Class. Quant. Grav.}, 27:114006, 2010.

\bibitem{Buonanno:2006ui}
Alessandra Buonanno, Gregory~B. Cook, and Frans Pretorius.
\newblock {Inspiral, merger and ring-down of equal-mass black-hole binaries}.
\newblock {\em Phys. Rev.}, D75:124018, 2007.

\bibitem{Baker:2006ha}
John~G. Baker, James~R. van Meter, Sean~T. McWilliams, Joan Centrella, and
  Bernard~J. Kelly.
\newblock {Consistency of post-Newtonian waveforms with numerical relativity}.
\newblock {\em Phys. Rev. Lett.}, 99:181101, 2007.

\bibitem{Pan:2007nw}
Yi~Pan et~al.
\newblock A data-analysis driven comparison of analytic and numerical
  coalescing binary waveforms: Nonspinning case.
\newblock {\em Phys. Rev.}, D77:024014, 2008.

\bibitem{Buonanno:2007pf}
Alessandra Buonanno et~al.
\newblock Toward faithful templates for non-spinning binary black holes using
  the effective-one-body approach.
\newblock {\em Phys. Rev.}, D76:104049, 2007.

\bibitem{Hannam:2007ik}
Mark Hannam, Sascha Husa, Ulrich Sperhake, Bernd Brugmann, and Jose~A.
  Gonzalez.
\newblock {Where post-Newtonian and numerical-relativity waveforms meet}.
\newblock {\em Phys. Rev.}, D77:044020, 2008.

\bibitem{Hannam:2007wf}
Mark Hannam, Sascha Husa, Bernd Bruegmann, and Achamveedu Gopakumar.
\newblock {Comparison between numerical-relativity and post-Newtonian waveforms
  from spinning binaries: the orbital hang-up case}.
\newblock {\em Phys. Rev.}, D78:104007, 2008.

\bibitem{Gopakumar:2007vh}
Achamveedu Gopakumar, Mark Hannam, Sascha Husa, and Bernd Bruegmann.
\newblock {Comparison between numerical relativity and a new class of
  post-Newtonian gravitational-wave phase evolutions: the non-spinning
  equal-mass case}.
\newblock {\em Phys. Rev.}, D78:064026, 2008.

\bibitem{Hinder:2008kv}
Ian Hinder, Frank Herrmann, Pablo Laguna, and Deirdre Shoemaker.
\newblock {Comparisons of eccentric binary black hole simulations with
  post-Newtonian models}.
\newblock {\em Phys. Rev.}, D82:024033, 2010.

\bibitem{Lousto:2010ut}
Carlos~O. Lousto and Yosef Zlochower.
\newblock {Extreme-Mass-Ratio-Black-Hole-Binary Evolutions with Numerical
  Relativity}.
\newblock 2010.

\bibitem{Campanelli:2007cg}
Manuela Campanelli, Carlos~O. Lousto, Yosef Zlochower, and David Merritt.
\newblock Maximum gravitational recoil.
\newblock {\em Phys. Rev. Lett.}, 98:231102, 2007.

\bibitem{Lousto:2007db}
Carlos~O. Lousto and Yosef Zlochower.
\newblock {Further insight into gravitational recoil}.
\newblock {\em Phys. Rev.}, D77:044028, 2008.

\bibitem{vanMeter:2010md}
James~R. van Meter, M.~Coleman Miller, John~G. Baker, William~D. Boggs, and
  Bernard~J. Kelly.
\newblock {A General Formula for Black Hole Gravitational Wave Kicks}.
\newblock 2010.

\bibitem{Brandt97b}
S.~Brandt and B.~Br{\"u}gmann.
\newblock A simple construction of initial data for multiple black holes.
\newblock {\em Phys. Rev. Lett.}, 78(19):3606--3609, 1997.

\bibitem{Ansorg:2004ds}
Marcus Ansorg, Bernd Br\"ugmann, and Wolfgang Tichy.
\newblock A single-domain spectral method for black hole puncture data.
\newblock {\em Phys. Rev. D}, 70:064011, 2004.

\bibitem{Zlochower:2005bj}
Y.~Zlochower, J.~G. Baker, M.~Campanelli, and C.~O. Lousto.
\newblock Accurate black hole evolutions by fourth-order numerical relativity.
\newblock {\em Phys. Rev. D}, 72:024021, 2005.

\bibitem{Lousto:2007rj}
Carlos~O. Lousto and Yosef Zlochower.
\newblock {Foundations of multiple black hole evolutions}.
\newblock {\em Phys. Rev.}, D77:024034, 2008.

\bibitem{Schnetter-etal-03b}
Erik Schnetter, Scott~H. Hawley, and Ian Hawke.
\newblock Evolutions in {3D} numerical relativity using fixed mesh refinement.
\newblock {\em Class. Quantum Grav.}, 21(6):1465--1488, 21 March 2004.

\bibitem{Thornburg2003:AH-finding}
Jonathan Thornburg.
\newblock A fast apparent-horizon finder for 3-dimensional {C}artesian grids in
  numerical relativity.
\newblock {\em Class. Quantum Grav.}, 21(2):743--766, 21 January 2004.

\bibitem{Dreyer02a}
Olaf Dreyer, Badri Krishnan, Deirdre Shoemaker, and Erik Schnetter.
\newblock Introduction to {Isolated} {Horizons} in {Numerical} {Relativity}.
\newblock {\em Phys. Rev. D}, 67:024018, 2003.

\bibitem{Campanelli99}
M.~Campanelli and C.~O. Lousto.
\newblock Second order gauge invariant gravitational perturbations of a {K}err
  black hole.
\newblock {\em Phys. Rev. D}, 59:124022, 1999.

\bibitem{Lousto:2007mh}
Carlos~O. Lousto and Yosef Zlochower.
\newblock A practical formula for the radiated angular momentum.
\newblock {\em Phys. Rev. D}, 76:041502(R), 2007.

\bibitem{Campanelli:2006gf}
M~Campanelli, C~O Lousto, and Y~Zlochower.
\newblock Last orbit of binary black holes.
\newblock {\em Phys. Rev. D}, 73:061501(R), 2006.

\bibitem{Boyle:2007ft}
Michael Boyle et~al.
\newblock High-accuracy comparison of numerical relativity simulations with
  post-newtonian expansions.
\newblock {\em Phys. Rev.}, D76:124038, 2007.

\bibitem{Lousto:2010xk}
Carlos~O. Lousto and Yosef Zlochower.
\newblock {Modeling maximum astrophysical gravitational recoil velocities}.
\newblock 2010.

\bibitem{Lousto:2009ka}
Carlos~O. Lousto, Hiroyuki Nakano, Yosef Zlochower, and Manuela Campanelli.
\newblock {Statistical studies of Spinning Black-Hole Binaries}.
\newblock {\em Phys. Rev.}, D81:084023, 2010.

\bibitem{Anderson:2009fa}
Matthew Anderson, Luis Lehner, Miguel Megevand, and David Neilsen.
\newblock {Post-merger electromagnetic emissions from disks perturbed by binary
  black holes}.
\newblock {\em Phys. Rev.}, D81:044004, 2010.

\end{thebibliography}

\end{document}